\documentclass{PoS}
\usepackage{amsmath} \usepackage{graphicx} \usepackage{amsfonts}
\usepackage{array} \usepackage{amsthm} \usepackage{bm}

\usepackage{latexsym}
\newcommand{\e}{{\rm e}}

\newcommand{\bw}{\begin{widetext}}
\newcommand{\ew}{\end{widetext}}
\newcommand{\be}{\begin{equation}}
\newcommand{\ee}{\end{equation}}
\newcommand{\bea}{\begin{eqnarray}}
\newcommand{\eea}{\end{eqnarray}}
\newcommand{\nn}{\nonumber}

\title{Hidden symmetries in 5D supergravities and black rings}

\ShortTitle{Hidden symmetries in 5D supergravities}

\author{\speaker{Dmitry V. Gal'tsov} \\
        Moscow State University\\
        E-mail: \email{galtsov@phys.msu.ru}}

\author{Nikolai G. Scherbluk  \\
        Moscow State University\\
        E-mail: \email{shcherbluck@mail.ru}}

\abstract{We construct generating technique for 5D minimal and
$U(1)^3$ supergravities based on hidden symmetries arising in
dimensional reduction to three dimensions. In the three-vector case
the symmetry is $SO(4,4)$, and the minimal case corresponds to
contraction of this group to $G_{2(2)}$. The matrix representation
is presented applicable to both cases and the generating
transformations preserving an asymptotic structure are listed. Our
transformations contain enough free parameters to construct the
general charged black ring in $U(1)^3$ theory starting with known
solutions. To avoid a complicated inverse dualisation in the
component form we introduce the matrix-valued dualisation which
opens the way to derive new solutions purely algebraically from the
coset representation of the seed.}

\FullConference{Black Holes in General Relativity and String Theory\\
         August 24-30 2008\\
         Veli Losinj,Croatia}

\begin{document}

\section{Introduction}
Recent interest to black rings \cite{empare} stimulated search of
new generating techniques for five-dimensional Einstein equations
both vacuum and coupled to vector and/or scalar fields. An efficient
tool to proliferate exact solutions to D-dimensional Einstein
equations depending on three coordinates consists in dimensional
reduction based on the assumption of existence of $D-3$ commuting
Killing symmetries (toroidal reduction). Starting with D-dimensional
Einstein equations coupled to scalar and vector fields one is able
to derive a three-dimensional gravity coupled sigma-model in which
the target space variables incorporate the initial scalars, vectors
and moduli of the toroidal reduction. For a particular class of
theories the target space turns out to be a coset space  $G/H$ where
$G$ is some semi-simple group known as the hidden symmetry group
(for a recent review see \cite{gerard}). This symmetry can be used
to generate new solution form known ones with the same
three-dimensional metric. The sigma-model representation may also
serve a basis of further reduction to two dimensions \cite{ga}
(looking for solutions depending only on two variables), where more
powerful methods can be developed such as inverse scattering
technique \cite{BeZa}. Recently such an approach to vacuum
five-dimensional relativity has undergone an impressive development
\cite{Har} and resulted in construction of rather sophisticated ring
configurations \cite{inve}. For charged rings no such technique was
available so far, though generation via some restricted
transformation involving vector fields were used \cite{yaza}. Our
matrix formulation opens a way to develop such methods in the
general case of minimal and $U(1)^3$ 5D supergravities describing
charged configurations.

Sigma-model generating technique for minimal five-dimensional
supergravity was developed in \cite{bccgsw,clem}, for an earlier
discussion of hidden symmetries in this theory see \cite{mizo}. The
hidden symmetry is this case is the non-compact version $G_{2(2)}$
of the lowest exceptional group $G_2$. To formulate the solution
generating technique one has to use some matrix representation of
the coset. Representing the seed solution in the matrix terms and
acting by   symmetry transformations one can extract the sigma-model
variables for  new solutions. In \cite{bccgsw,clem} an explicit
$7\times 7$ representation of the coset $G_{2(2)}/SL(2,R)^2$ was
constructed using the representation of $G_2$  found  by Gunyadin
and Gursey \cite{gugu}. The generalization to the case of
five-dimensional supergravity with three $U(1)$ vector and two
scalar fields was given in \cite{GS}. Apart from being more general,
this theory is interesting by the fact that the corresponding hidden
symmetry is given by  a familiar group $SO(4,4)$. Actually, one of
the ways to construct the matrix representation of $G_{2(2)}$
consists in using $SO(4,4)$ as a starting point \cite{gugu} and
imposing suitable constraints.  The matrix representation of the
relevant three-dimensional coset $SO(4,4)/SO(4)\times SO(4)$ is
given  in terms of the $8\times 8$ matrices which are split into the
$4\times 4$ blocks. By freezing the scalar moduli  and identifying
the vector fields  one reduces this theory  to minimal 5D
supergravity thus providing an alternative formulation of the
technique of \cite{bccgsw}  in terms of the $8\times 8$ matrices.

Generation of the new solution  appeals to transformation  of the
target space variables by the hidden symmetry group. This part of
the procedure is purely algebraic. Another part consists in solving
the differential dualisation equations relating the target space
variable to the metric and vector fields \cite{bccgsw,GS}. These
equations have to be solved twice: first for the seed solution to
obtain its description in terms of the coset matrix, and then for
the transformed solution in order to extract the metric and the
matter fields from the transformed coset matrix. Solving these
equations may present technical difficulties on the second step if
the generating transformations are complicated enough. To remedy
this problem, we propose to pass to dualized variables in the matrix
form. Such a possibility is suggested by the fact that the
three-dimensional dual to the sigma-model matrix-valued current
one-form is closed by virtue of the equations of motion. Then
locally it is exact, and this provides the matrix-valued one-form
whose exterior derivative is dual to the the initial matrix current.
This dual matrix transforms under the global action of the hidden
symmetry by some related transformation, and thus it can be found
algebraically from the corresponding seed matrix. From this can read
out the metric and matter fields of the transformed solution
algebraically avoiding the inverse dualisation problem.

Our primary motivation to develop the generating techniques for 5D
supergravity was lack of the general black ring solution possessing
the electric charges (one in the minimal case and three  in the
$U(1)^3$ case), the magnetic charges, the mass and two independent
rotation parameters \cite{empare}. In \cite{bccgsw} an attempt was
made to construct a charged black ring starting with the neutral
solution with two rotation parameters found by Pomeranski and Senkov
\cite{posen}. But the resulting solution was plagued with a conical
singularity. To be able to derive a regular solution, one has to
start with a non-regular seed solution with an extra free parameter,
which can be fixed after the transformation. In principle, from the
counting of free parameters in the transformations preserving
asymptotic behavior of black rings, one finds that the general black
ring can be generated indeed starting from some known solutions. But
so far all attempts to find such a solution in a concise form were
unsuccessful.

In this paper we illustrate the application of our technique
generating new Kaluza-Klein squashed black holes. These black holes
look as five-dimensional near the event horizon exhibiting the $S^3$
strucure, but asymptotically $S^3$ collapses to a twisted bundle of
$S^1$ over  $S^2 $ with a constant radius of $S^1$ and growing
radius of $S^2$. Thus at infinity they become four-dimensional
objects  with a compactified fifth dimension. One such solution to
five-dimensional Einstein-Maxwell system was proposed by Ishihara
and Matsuno \cite{IM} (non-rotating). Its physical parameters and
thermodynamical properties were investigated in \cite{Ca,Kuri}.  A
certain class (but not all) of squashed black holes can be obtained
by the so-called squashing transformation. This procedure was
applied to asymptotically flat \cite{IM,wang,NIMT} and
non-asymptotically flat solutions such as Kerr-G\"{o}del black holes
\cite{TIMN,MINT,TI}. In an attempt to enlarge the class of
solutions, more recently Tomizawa, Yasui and Morisawa \cite{tym}
applied $G_{2(2)}$ transformations of \cite{bccgsw} to construct a
generalization of the charged Rasheed black hole \cite{rash}
obtaining  a new solution with four independent parameters: mass,
angular momentum, Kaluza-Klein parameter $\beta$ (in the notation of
\cite{rash}) and an electric charge. Here we will derive a more
general five-parametric solution adding as an independent parameter
the quantity $\alpha$ of \cite{rash}, which corresponds to an
electric charge in the four-dimensional interpretation of the
Rasheed solution.

\section{General setting}

The $U(1)^3$ 5D supergravity  may be regarded as a truncated
toroidal compactification of the 11D supergravity:
 \be \label{ans11} I_{11} = \frac{1}{16\pi
G_{11}}\int\left(R_{11}\star_{11} \mathrm{1}-\frac12 F_{[4]}\wedge
\star_{11} F_{[4]} - \frac16F_{[4]}\wedge F_{[4]} \wedge
A_{[3]}\right),
 \ee
where $ F_{[4]} = dA_{[3]}$. Assuming an ansatz for the metric \be
ds_{11}^2 = ds^2_{5} + X^1 \left( dz_1^2 + dz_2^2 \right) + X^2
\left( dz_3^2 + dz_4^2 \right) + X^3 \left( dz_5^2 + dz_6^2
\right),\label{ds_11}\ee  and the form field  $$ A_{[3]} = A^1
\wedge dz_1 \wedge dz_2 + A^2 \wedge dz_3 \wedge dz_4 + A^3 \wedge
dz_5 \wedge dz_6,
$$ where all functions are independent of $z$, we obtain the
 the bosonic sector of 5D supergravity  coupled to three scalar
moduli $X^I$ ($I=1,2,3$), satisfying the constraint $X^1X^2X^3=1$,
and to three vector fields $A^I$:
 \bea\label{L5}
I_5 \!&=&\! \frac{1}{16 \pi G_5} \int\!\! \left( R_5 \star_5 1 \!-\!
\frac12 G_{IJ} dX^I \!\wedge\! \star_5 dX^J \!-\! \frac12G_{IJ} F^I
\!\wedge\! \star_5 F^J \!-\!
\frac{1}{6} \delta_{IJK} F^I \!\wedge\! F^J \!\wedge\! A^K \right)\!\!, \\
G_{IJ}&=&{\rm diag}\left((X^1)^{-2},\ (X^2)^{-2},\
(X^3)^{-2}\right),\quad F^{I}=dA^{I},\quad I,J,K=1,2,3.\nn
 \eea
Here the Chern-Simons coefficients $\delta_{IJK}=1$ for the indices
$ I,J,K $ being a permutation of 1, 2, 3, and zero otherwise.
Contraction of the above theory to minimal $5D$ supergravity is
effected via an identification of the vector fields:
$$A^1=A^2=A^3=\frac{1}{\sqrt{3}}A,$$ and freezing out  the moduli:
$X^1=X^2=X^3=1$. This leads to the Lagrangian  \be \mathcal{L}_5 =
R_5 \star_5 \mathbf{1} - \frac12 F\wedge \star_5 F -
\frac{1}{3\sqrt{3}} F\wedge F \wedge A.\nn \ee It is worth noting
that the 5D Einstein-Maxwell theory without the Chern-Simons term
does not lead to the three-dimensional sigma model with  a
semi-simple hidden symmetry group, so in this case the solution
generating technique can be formulated only for the static
truncation of the theory. This explains why the charged rotating
black hole solutions  are not known analytically.

\subsection{Four-dimensional view}Consider reduction of the D=5 action (\ref{L5}) to  four
dimensions. We assume that the   5D space-time has the structure
${\cal M}_5={\cal M}_4 \times S^1$, where $S^1$ is a circle, and is
parameterized by the coordinates $\{x^{\mu},z\},\ \mu=1,\ldots, 4$
with $z$ relating to the circle. Following to the standard procedure
we decompose the 5D metric as
 \be
  ds_5^2=e^{\frac{\phi}{\sqrt3}}ds_4^2+e^{-\frac{2\phi}{\sqrt3}}(dz+a)^2,
  \ee
where the is  $ds_4^2=g_{\mu\nu}(x)dx^{\mu}dx^{\nu}$,  the
Kaluza-Klein one-form is $a=a_{\mu}dx^{\mu}$ and $\phi$ is the
dilaton.
 In a similar way the 5D vector fields $A^I(x^{\mu},z)$ are
 decomposed as
 \be
   A^I(x^{\mu},z)=A^I(x^{\mu})+u^I dz,
  \ee
where $u^I$ are the  axions. All the above fields do not  depend on
$z$.
 Inserting these decompositions into the 5D action we get the 4D
lagrangian
 \bea\label{L4}
  {\cal
  L}_4&=&R_4\star 1-\frac12\star d\phi\wedge d\phi-\frac12G_{IJ}\star dX^I\wedge dX^J
  -\frac12
  e^{\frac{2\phi}{\sqrt3}}G_{IJ}\star du^I\wedge
  du^J-\frac12e^{-\sqrt3\phi}\star{\cal
  F}\wedge {\cal F}\\
  &-&\frac12e^{-\frac{\phi}{\sqrt3}}G_{IJ}\star F^I
  \wedge F^J-\frac12\delta_{IJK}dA^I\wedge dA^J u^K,\nn
  \eea
where $ {\cal F}=da$ and $F^I=dA^I-du^I\wedge a$ are the field
strength two-forms. Our purpose is to rewrite this lagrangian in the
form exhibiting the S-duality symmetry. First of all we consider the
scalar part of (\ref{L4}) written in the following form
 $$
  e_4^{-1}{\cal L}_{scal}=\frac12\left((\partial\phi)^2+
  G_{IJ}\partial X^I\partial X^J+e^{\frac{2\phi}{\sqrt3}}G_{IJ}\partial u^I\partial
  u^J\right)=\hat{{\cal
  G}}_{AB}(\hat{\Phi})\partial\hat{\Phi}^A\partial\hat{\Phi}^B,\quad
  A,B=1,\ldots,
  6,
  $$
 where $\partial\equiv\partial/\partial x^{\mu}$, $e_4$ is the Hodge
 dual to unity: $e_4\equiv\star
1=\sqrt{-g}\,d^4 x$ and all index operations refer to the metric
$g_{\mu\nu}$. The potentials $\hat{\Phi}^A$ combine the six
variables
 $\{X^1,X^2,\phi,u^I\}$ and realize the map $\hat{\Phi}^A:\ x^{\mu}\in {\cal M}_4\
 \rightarrow\ \hat{\Phi}^A(x^{\mu})\in {\cal M}_{scal}$ between the 4D Minkowskian
 space-time and the target space with the metric $\hat{{\cal
  G}}_{AB}(\hat{\Phi})$. Replacing the dilaton
  $\phi$ and the moduli $X^I$ by the new variables $\alpha^I$: $\alpha^I=
  \phi/\sqrt{3}-\ln X^I$
  enable us to simplify ${\cal L}_{scal}$ as follows
  $$
   e_4^{-1}{\cal L}_{scal}=\frac12\sum_I
   \left((\partial\alpha^I)^2+e^{2\alpha^I}(\partial
   u^I)^2\right).
   $$
 The structure of the scalar manifold ${\cal M}_{scal}$ becomes more
transparent in terms of three complex potentials
 $z^I=u^I+ie^{-\alpha^I}$: $$e_4^{-1}{\cal L}_{scal}=\frac12\sum |\partial z^I|^2/(\mathrm{Im}\
 z^I)^2.$$ The lagrangian ${\cal L}=\frac12|\partial z|^2/(\mathrm{Im}\
 z)^2$ invariant under the group $SL(2,\mathbb{R})$
and the corresponding target space metric is the K\"ahler space
$SL(2,\mathbb{R})/SO(2)$. So in our case the isometry group of
${\cal M}_{scal}$ is $\hat G=(SL(2,\mathbb{R}))^3$ and the
corresponding target space is $\hat G/\hat H={\cal
M}_{scal}=(SL(2,\mathbb{R})/SO(2))^3$ with the metric
 $$
 \hat{{\cal
  G}}_{AB}(\hat{\Phi})d\hat{\Phi}^A d\hat{\Phi}^B=
 \frac12\left((d\phi)^2+
  G_{IJ}d X^I dX^J+e^{\frac{2\phi}{\sqrt3}}G_{IJ}d u^I
  du^J\right)=\frac12\sum |dz^I|^2/(\mathrm{Im}\
 z^I)^2.
 $$

As the second step, we reformulate the vector part of the lagrangian
(\ref{L4}) according with the structure of the bosonic lagrangian of
$N=2$ supergravity coupled to  vector multiplets (for a review see
 the Ref.\cite{Fre}). We express it in terms of the field two-forms
$\widetilde F^I$ and ${\cal F}$ obeying to the Bianchi identities
$d\widetilde F^I=0$ and $d{\cal F}=0$ respectively. To extract the
two-forms $\widetilde F^I$ one has to combine the exterior
derivative $d(u^Ia)$ in $F^I=dA^I-du^I\wedge a$. As result we have
$F^I=\widetilde F^I+u^I{\cal F}$, where $\widetilde F^I=d\widetilde
A\equiv d(A^I-u^Ia)$. Inserting the two-forms $F^I$ and $dA^I$
expressed via $\widetilde F$ and ${\cal F}$ into (\ref{L4}) and
integrating by parts the terms  $\delta_{IJK}\widetilde F^I\wedge
du^J u^K\wedge a$ and $\delta_{IJK}du^I u^J u^K\wedge a\wedge {\cal
F}$ we will obtain for the vector part of the 4D lagrangian
 \bea\label{L_tens}
  {\cal L}_{vect}&=&\frac12\Bigl[e^{-\sqrt3\phi}\star{\cal F}\wedge{\cal F}+e^{-\frac{\phi}{\sqrt3}}
  G_{IJ}\left(\star \widetilde F^I\wedge \widetilde F^J+2\star{\cal F}u^{[I}\wedge \widetilde F^{J]}
  +u^I u^J\star{\cal F}\wedge{\cal F}\right)  \\
  &+& \delta_{IJK}\Bigl(\widetilde F^I\wedge \widetilde F^J u^K+\widetilde F^I u^J u^K\wedge{\cal F}
  +\frac13 u^I u^J u^K {\cal F}\wedge {\cal F}\Bigr) \Bigl].\nn
  \eea
Denote the field strength  and its Hodge dual as
 ${\cal F}=\frac12 {\cal F}_{\mu\nu}dx^{\mu}\wedge dx^{\nu}$ and
 $\star{\cal F}_{\mu\nu}=\frac12{\cal F}^{\alpha\beta
}\epsilon_{\alpha\beta\mu\nu}$,
 where $\epsilon_{\alpha\beta\mu\nu}$ is the totally antisymmetric
Levi-Civita tensor with $(-g)^{1/2}$. Assuming that $dx^{\mu}\wedge
dx^{\nu}\wedge dx^{\alpha}\wedge
dx^{\beta}=-\epsilon^{\mu\nu\alpha\beta}e_4,$ we find
$$
 \star{\cal F}\wedge {\cal F}=\frac12 {\cal F}_{\mu\nu}{\cal
 F}^{\mu\nu}e_4=\frac12{\cal F}^2e_4,\quad
 {\cal F}\wedge {\cal F}=-\frac12 {\cal F}_{\mu\nu}\star{\cal
 F}^{\mu\nu}e_4=-\frac12{\cal F}\star{\cal F}e_4.
 $$
Note that in the 4D Lorentzian signature space  the double Hodge
dual is $\star\star=-1$. We then combine the field tensors
$\widetilde F^I_{\mu\nu}$ and ${\cal F}_{\mu\nu}$ into the 4-column
${\cal B}_{\mu\nu}=\left(\begin{array}{c}
  \widetilde F^I_{\mu\nu} \\
  {\cal F}_{\mu\nu} \\
\end{array}\right)$ and  rewrite (\ref{L_tens}) in the
matrix form adopted in \cite{BMG,BM}:
$$
 e_4^{-1}{\cal L}_{vect}=\frac14{\cal B}_{\alpha\beta}^T (\hat\mu {\cal B}^{\alpha\beta}-\frac{1}{\sqrt2}\hat\nu\star{\cal
 B}^{\alpha\beta}),
$$
 where the symmetric $4\times 4$ matrices $\hat\mu$ and $\hat\nu$ are given by
 $$
  \hat\mu=\left(%
\begin{array}{cc}
  e^{-\frac{\phi}{\sqrt3}}G_{IJ} & e^{-\frac{\phi}{\sqrt3}}G_{IJ}u^J \\
  e^{-\frac{\phi}{\sqrt3}}G_{IJ}u^J & G_{IJ}u^I u^J+e^{-\sqrt3\phi} \\
\end{array}%
\right),\quad \hat\nu=\sqrt2\left(%
\begin{array}{cc}
  \delta_{IJK}u^K & \frac12\delta_{IJK}u^J u^K\\
  \frac12\delta_{IJK}u^J u^K & 2u^1u^2u^3 \\
\end{array}%
\right).
  $$
 This lagrangian yields the field equations for  ${\cal
 B}_{\alpha\beta}^T$: $\nabla_{\alpha}(\hat\mu {\cal B}^{\alpha\beta}-\frac{1}{\sqrt2}\hat\nu\star{\cal
 B}^{\alpha\beta})=0$. Introducing the dual field strength ${\cal
 H}_{\alpha\beta}$ as $\star{\cal H}^{\alpha\beta}=\hat\mu {\cal B}^{\alpha\beta}-\frac{1}{\sqrt2}\hat\nu\star{\cal
 B}^{\alpha\beta}$ we see that the above equations are the Bianchi identities for
 ${\cal H}_{\alpha\beta}$. Therefore the lagrangian ${\cal L}_{vect}$ takes the form
 manifestly S-duality symmetric:
  $$
   e_4^{-1}{\cal L}_{vect}=\frac14{\cal B}_{\alpha\beta}^T\star{\cal
   H}^{\alpha\beta}=\frac18{\Im}^T\Sigma_1\star{\Im},\quad {\Im}=\left(\begin{array}{c}
     {\cal B_{\alpha\beta}} \\
     {\cal H_{\alpha\beta}} \\
   \end{array}\right),\quad \Sigma_1=\left(%
\begin{array}{cc}
  0 & 1 \\
  1 & 0 \\
\end{array}%
\right).
  $$
 It can be checked that relation between $\Im$ and $\star\Im$ is
 given by
 $$
  \Im=\Omega\, \hat P\star\Im,
 $$
where $\Omega=
  \left(%
\begin{array}{cc}
  0 & 1 \\
  -1 & 0 \\
\end{array}%
\right)$ is the $8\times 8$ symplectic metric and $\hat P$ is the
$8\times 8$ matrix depending on the potentials of the scalar
manifolds ${\cal M}_{scal}$
$$
 \hat P=\left(%
\begin{array}{cc}
  \hat\mu+\hat\nu\hat\mu^{-1}\hat\nu & \hat\nu\hat\mu^{-1} \\
  \hat\mu^{-1}\hat\nu & \hat\mu^{-1}  \\
\end{array}%
\right).
$$
The matrix $\hat P$ provides the representation $\gamma$ of the
coset element $\pi(\hat\Phi^A)$, namely $\gamma: \pi\in {\cal
M}_{scal}\ \rightarrow\ \gamma(\pi)=\hat P$. We then have
$$\hat{{\cal
  G}}_{AB}d\hat{\Phi}^A d\hat{\Phi}^B=-\frac{1}{16}\textrm{Tr}(d\hat P\,d\hat P^{-1})=
  -\frac18\textrm{Tr}(d\mu d\mu^{-1}-d\nu\mu^{-1}d\nu\mu^{-1}).$$

Consider diffeomorphism $\hat\Phi^A \rightarrow \hat\Phi^{A}{'}$,
which leave invariant the target space metric. It corresponds to the
action of some element $\hat g$ belonging to the isometry group of
the target space $\hat g\in \hat G$. In terms of the matrix
representation $\gamma$ this means that the coset matrix $\hat
R\equiv \Sigma_1\hat P$ transforms as $\hat R \rightarrow \hat
R'=\gamma(\hat g)\hat R\gamma(\hat g^{-1})$. Inserting the
expression $\star \Im=-\Omega\hat P\Im$ into the ${\cal L}_{tens}$
and keeping in mind that $\Sigma_1\Omega=-\Omega\Sigma_1$ we will
obtain for the tensor part of the lagrangian:
$$e_4^{-1}{\cal L}_{vect}=\frac18{\Im}^T\Omega\hat R\Im.$$
 If we now demand   this lagrangian to be invariant under the
 action of $\gamma(\hat g)$, we get the restrictions for the
 element $\widetilde g\in \hat G$ acting on the column as $\Im \rightarrow \gamma(\widetilde
 g)\Im$. Performing the transformation we have
 $${\cal L}_{vect} \rightarrow {\cal L}_{vect}'=\frac18\Im^T\gamma(\widetilde
 g)^T
 \Omega\gamma(\hat g)\hat R\gamma(\hat g^{-1})\gamma(\widetilde g)\Im.$$
 Thus the conditions for $\gamma(\widetilde g)$ are  $\gamma(\widetilde g)=\gamma(\hat g)$
 and $\gamma(\hat g)^T\Omega\gamma(\hat g)=\Omega$. This relation
 means that there is the symplectic embedding of the isometry
 group into the symplectic group $\hat G \rightarrow Sp(8,\mathbb{R})$ \cite{abcaffm}.
In other words, $\gamma(\hat g)$ provides
 the symplectic representation of $\hat g$ which rotates the
 fields $\Im$. Note the full 4D lagrangian can be written in the
 following form
 $$
  e_4^{-1}{\cal L}_4=R_4+\frac{1}{16}\textrm{Tr}(\partial\hat R\,\partial\hat R^{-1})-\frac18\Im^T\Omega\hat R\Im.
 $$
 Thus the S-duality group for the four-dimensional reduction of the
 $U(1)^3$ supergravity is $SL(2,R)^3$, reducing to $SL(2,R)$
in the minimal case.

\section{3D sigma-model} Consider now further
reduction to three dimensions. It is convenient to restart from 11D
supergravity. An overall assumption for the 11D manifold will be
${\cal M}_{11}=T^6\times\Sigma\times{\cal M}_3 $ where $\Sigma$ is
$T^2$ if both  these Killing vectors are asymptotically space-like,
or $T^1\times \mathbb{R}$ if one of them is asymptotically
time-like. The full set of 11D coordinates $x^{N},\ N=1,\ldots,11$
is thus split into $z^a\in T^6,\ a=1,\ldots,6$, $x^i\in {\cal M}_3,\
i=1,\ldots,3$ and $z^p\in \Sigma,\ p=7,8$. The decomposition of the
5D metric  is given by
 \be
  ds_5^2=\lambda_{pq}(dz^p+a^p)(dz^q+a^q)-\kappa\tau^{-1}h_{ij}dx^i
 dx^j,\label{ds_5}
 \ee
where all metric functions are independent on $z^a$ and $z^p$.  The
5D metric components are parameterized by the KK one-forms
$a^p=a^p_idx^i$, the three-dimensional metric $h_{ij}$ of  ${\cal
M}_3$ and the scalars $\varphi_1,\varphi_2,\chi$, which are arranged
in the following $2\times 2$ matrix
 $$
 \lambda=\e^{-\frac{2}{\sqrt3}\varphi_1}\left( \begin{array}{cc}
1&
\chi \\
\chi & \chi^2+\kappa \e^{\sqrt{3}\varphi_1-\varphi_2}\end{array}
\right),\qquad  \det\lambda\equiv -\tau=\kappa
 \e^{-\frac{1}{\sqrt3}\varphi_1-\varphi_2}, \label{def_lambda}
 $$
where  $\kappa=\pm$ is responsible for the signature: $\kappa=1$ for
space-like $z^8$ , and $\kappa=-1$ for time-like $z^8$. The ansatz
(\ref{ds_11}) leads to the five-dimensional action (\ref{L5}). The
 5D $U(1)$ gauge fields $A^I$ reduce to the 3D one-forms
 $A^I(x^i)$ and the six axions collectively denoted as the
2D-covariant doublet
 $\psi_p^I=(u^I,v^I)$ with the index $p$ relative to the metric
 $\lambda_{pq}$
 $$
  A^I(x^i,z^7,z^8)=A^I(x^i)+\psi_p^I dz^p=A^I(x^i)+u^I dz^7+v^I
  dz^8.
 $$

To obtain the three-dimensional sigma-model one has to dualize the
electro-magnetic (EM) one-forms $A^I$ and the KK one-forms $a^p$ to
scalars, which will be denoted as  $\mu_I$ and $\omega_p$. The
 dualisation equations read:
 \bea
  && \tau\lambda_{pq}da^q=\star V_p,\nn\\
  dA^I&=&d\psi_q^I\wedge a^q+\tau^{-1}G^{IJ}\star
  G_J,\label{eqs_of_dual}
 \eea
where the one-forms $G_I$ and $V_p$ are given by
  \bea
  &&G_I=d\mu_I+\frac12\delta_{IJK}d\psi_p^J
 \psi_q^K\varepsilon^{pq},\nn\\
  && V_p=d \omega_p- \psi_p^I  \Big( d \mu_I+\frac16 \delta_{IJK}d\psi_q^J
 \psi_r^K\varepsilon^{qr}\Big).\nn
  \eea
In the component form the Eqs.(\ref{eqs_of_dual}) read:\footnote{the
antisymmetrization is assumed with 1/2.}
 \bea
 &&\lambda_{pq}\partial^{{[}i} a^{j{]q}}=\frac{1}{2\tau \sqrt
 h}\varepsilon^{ijk}\Bigg[\partial_k \omega_p- \psi_p^I  \left(\partial_k \mu_I+
 \frac16 \delta_{IJK}\partial_k\psi_r^J
 \psi_t^K\varepsilon^{rt}\right)\Bigg],\nn \\
 &&\partial^{{[}i} A^{j{]}I}=a^{q{[}j}\partial^{i{]}}
 \psi_q^I+\frac{1}{2\tau\sqrt{h}}\varepsilon^{ijk}G^{IJ}\left(\partial_k\mu_J+
 \frac12\delta_{JKL}\partial_r\psi_p^K
 \psi_q^L\varepsilon^{pq}\right). \label{dualizequ}
 \eea
Substituting the   metric $ds_5^2$ in the form  (\ref{ds_5}) into
the 5D action (\ref{L5}) and performing   dualisation via
Eqs.(\ref{eqs_of_dual})  one derive the 3D gravity coupled
sigma-model:
 \be
I_3=\frac{1}{16\pi G_3}\int \sqrt{|h|}\left(R_3-{\cal
G}_{AB}\frac{\partial\Phi^A}{\partial
x^i}\frac{\partial\Phi^B}{\partial x^j}h^{ij}\right)d^3x,\label{L3}
 \ee
where the Ricci scalar $R_3$ is build using the 3-dimensional metric
$h_{ij}$. The set of potentials \footnote{The set
$\vec\phi=(\phi_1,\phi_2,\phi_3,\phi_4)$ comprises four scalars
related to  previously introduced  $\varphi_1,\varphi_2,\chi$ and
$X^I$ via
 \bea
  \phi_1&=&\frac{1}{\sqrt2}\left(-\ln(X^3)+\frac{1}{\sqrt3}\varphi_1+\varphi_2\right),\quad
   \phi_2=\frac{1}{\sqrt2}\left(\ln(X^3)-\frac{1}{\sqrt3}\varphi_1+\varphi_2\right),\nn
 \\
 \phi_3&=&\frac{1}{\sqrt2}\left(\ln(X^3)+\frac{2}{\sqrt3}\varphi_1\right),\qquad\qquad
 \phi_4=\frac{1}{\sqrt2}\,\ln\frac{X^1}{X^2}.\nn
 \eea} $\Phi^A=(\vec\phi,  \psi^I,
\mu_I,\chi,\omega_p),$ $A,B=1,\ldots,16$ realizes the harmonic map
$\Phi^A:\ x^i\in {\cal M}_3\ \rightarrow\ \Phi^A(x^i)\in {\cal
M}_{scal}$ between the 3D space-time ${\cal M}_3$ and the target
space  ${\cal M}_{scal}$ with the metric ${\cal G}_{AB}(\Phi^C)$.
The target space line element $dl^2={\cal{G}}_{AB}d\Phi^Ad\Phi^B$
has the form
 \bea
 dl^2& =& \frac12 G_{IJ}(dX^IdX^J+d{{\psi^I}^T}\lambda^{-1}
d\psi^J)-\frac12\tau^{-1}G^{IJ}G_IG_J+\frac14 \mathrm{Tr} \left(
\lambda^{-1} d\lambda \lambda^{-1} d\lambda \right)\nn\\
 & +& \frac14\tau^{-2} d\tau^2 - \frac12\tau^{-1} V^T \lambda^{-1}
V.\label{TS_metric}
 \eea
It is invariant under the action of the 28-parametric isometry group
$SO(4,4)$. The target space manifold ${\cal M}_{scal}$ is isomorphic
to the coset ${\cal M}=SO(4,4)/H$, where the isotropy group $H$ is
$SO(4)\times SO(4)$ for $\kappa=1$ and  $SO(2,2)\times SO(2,2)$ for
$\kappa=-1$. That is there is an isomorphic map $\pi$:
$\Phi^A\rightarrow \pi(\Phi^A)\in {\cal M}$. Moreover if $g\in
SO(4,4)$ is some constant element of the isometry group then the
following transformations
$$
 \pi \rightarrow \pi'=g\circ\pi,\quad ds^2_3\rightarrow ds^2_3
$$
leave invariant the action (\ref{L3}).

As a convenient representative of the  coset $\pi(\Phi^A)\in {\cal
M}$ one can choose the matrix representation $\gamma: \pi
\rightarrow \gamma(\pi)\equiv {\cal V}$, where ${\cal V}$ is the
upper triangular matrix. We assume that ${\cal V}$ transforms under
the global action of the symmetry group $SO(4,4)$ by the right
multiplication and under the local action of the isotropy group $H$
by the left multiplication: ${\cal V}\to {\cal V}'=h(\Phi){\cal
V}g$, where $g$ and $h$ belong to the matrix representation $\gamma$
of $SO(4,4)$ and $H$ respectively. Given this representative, one
can construct the $H-$invariant matrix (which we denote the same
symbol ${\cal M}$ as the coset space)
 $$
 {\cal M}={\cal V}^T K {\cal V},
 $$
where $K$ is an involution matrix invariant under $H$: $h(\Phi)^T K
h(\Phi)=K,\label{h_K_h}$ and dependent on the coset signature
parameter $\kappa$. Then the transformation of the matrix ${\cal M}$
under $SO(4,4)$ will be
 \be
 {\cal M}\to {\cal M}'=g^T {\cal M}g.\label{trans_M}
 \ee
The target space metric (\ref{TS_metric}) in terms of the matrix
${\cal M}$ will read  \be
 dl^2=-\frac18 \mathrm{Tr}(d{\cal M}d{\cal M}^{-1}).\label{TS_by_M}
 \ee
Choosing suitable $8\times 8$ matrix representation $\gamma$ of the
isometry group $SO(4,4)$ we construct (see \cite{GS} for details )
the matrix representation of the coset ${\cal M}$ in terms of the
$4\times 4$ block matrices ${\cal P}={\cal P}^T$ and ${\cal
Q}=-{\cal Q}^{\widehat T}\footnote{$\widehat T$ denotes
transposition with respect to the minor diagonal}$ as follows
$$
 {\cal M}=\left( \begin {array}{cc}
 {\cal P}&{\cal P}{\cal Q}\\
 {\cal Q}^T{\cal P}&\widetilde{\cal P}+{\cal Q}^T{\cal P}{\cal Q}
 \end {array} \right),
 $$
 where the block matrices are given explicitly in the Appendix.
\subsection{Matrix dualisation}
  As we have discussed, the
dualisation equations  (\ref{dualizequ}) may present difficulties in
applications of the solution generating technique. We can improve
the situation performing dualisation in the matrix form. Introducing
the matrix-valued current one-form  ${\cal J}$
$$
{\cal J}={\cal J}_idx^i={\cal M}d{\cal M}^{-1}
$$
we can rewrite the 3-dimensional sigma-model action (\ref{L3}) in
the following form
$$
I_3=\frac{1}{16\pi G_3}\int \left(R_3\star 1-\frac18
\mathrm{Tr}({\cal J}\wedge\star{\cal J})\right).
$$
In this expression the Hodge dual $\star\ $ is assumed with respect
to the 3-dimensional metric $h_{ij}$. Variation of this action with
respect to ${\cal J}$ shows that the two-form $\star{\cal J}$ is
closed: \be d\star{\cal J}=0.\label{eqs_of_motion}\ee Variation with
respect to the metric leads to  three-dimensional Einstein
equations: \be (R_3)_{ij}=\frac18 \mathrm{Tr}({\cal J}_i{\cal
 J}_j). \ee
 The first equation (\ref{eqs_of_motion}) means that the
 matrix-valued
 two-forms $\star {\cal
 J}$ is locally exact, i.e., it can be  presented
 as the exterior derivative of some matrix-valued one-form ${\cal
 N}$, that is
 \be
  \star{\cal J}={\cal M}\star d{\cal M}^{-1}=d{\cal N}.\label{def_N}
 \ee
 The matrix ${\cal N}$ is defined up to  adding an arbitrary matrix-valued
  closed one-form, which can be determined by choosing suitable
 asymptotic conditions. Now comparing the matrix dualisation equation
 (\ref{def_N}) with the initial dualisation equations
 (\ref{dualizequ})
we find  the following purely algebraic relations between certain
 components of the matrix $({\cal N})_{ab},\ a,b=1,\ldots, 8$ are
and the previous variables $a^p$ and
  $A^I$, namely
 \bea
  && a^7=({\cal N})_{16},\quad a^8=({\cal N})_{17},\nn\\
  && A^{1}=\psi^1_{p}a^p+({\cal N})_{15},\quad A^{2}=\psi^2_{p}a^p+({\cal
  N})_{14},\quad A^{3}=\psi^3_{p}a^p-({\cal
  N})_{26}.\label{rel_for_N}
 \eea
 Thus, if one manages to find the matrix ${\cal N}$, the metric and
 matter fields can be extracted algebraically.

 For the following it is important that the definition (\ref{def_N}) and
 the transformation law for the matrix ${\cal M}$ (\ref{trans_M})
 under the global transformations $g\in SO(4,4)$ imply the
 following transformation of the matrix ${\cal N}$:
 $$
  {\cal N}\rightarrow {\cal N}\ '=g^T{\cal N}(g^T)^{-1}.
 $$
 Using (\ref{rel_for_N} one can read off the metric components and
 potentials without explicitly solving the differential dualisation
 equations.

\section{Solution generating technique}
The sigma-model presented in the previous sections gives rise to
 generating technique which allows to construct new solutions
 from the known ones.  Let the metric $h_{ij}$ and the set of
potentials $\Phi^A$ combined in the coset matrix ${\cal M}$
correspond to the metric   and the three-form  of some 11D seed
solution. One has to extract part of the target space potentials
from the seed solution algebraically and solve the differential
dualisation equations (\ref{eqs_of_dual}) to find the remaining
potentials. Using the action of the target space isometries one can
then construct a new solution of the sigma-model with the same
three-metric $h_{ij}'=h_{ij}$ and the coset matrix
$$
 {\cal M}'=g^T{\cal M}g\quad (\hbox{or }{\cal M}'=g{\cal
 M}g^T),\quad g\in SO(4,4).
$$
Note that  five target space variables
$\phi_1,\,\phi_2,\,\phi_3,\,\phi_4,\, \chi$ enter the
eleven-dimensional metric algebraically, via the moduli $X^I,\,
\lambda_{pq}$:
$$
  ds_{11}^2 = \sum_{I,a,a'} X^I \left( (dz^a)^2 + (dz^{a'})^2
  \right)+
  \lambda_{pq}(dz^p+a^p)(dz^q+a^q)+\tau^{-1}h_{ij}dx^idx^j,\ aa'=(12,34,56),
$$
while the KK vectors $a^p$ in the $T^2$ sector are related to the
target space potentials $\omega_p$ via dualisation. Similarly, in
the form-field sector,
$$
  A_{[3]}=(A^1+\psi_p^1 dz^p)\wedge dz^1 \wedge dz^2+
 (A^2+\psi_p^2 dz^p)\wedge dz^3 \wedge dz^4+
 (A^3+\psi_p^3 dz^p)\wedge dz^5 \wedge dz^6
$$
the six quantities $\psi_p^I$ are the target space potentials, while
the remaining one forms $A^I$ are related to the potentials $\mu_I$
via dualisation. So the set of transformed potentials
$\lambda_{pq}',\ (X^{I})'$ and $(\psi_p^{I})'$ can be explicitly
extracted from the coset matrix ${\cal M}'$. The remaining
components of the transformed metric $(ds_{11}^2)'$ and the 3-form
$(A_{[3]})'$ which are parametrized as the KK one-forms $(a^p)'$ and
the EM fields $(A^I)'$ are determined by the dualisation equations
(\ref{eqs_of_dual}). The inverse dualisation via the Eqs.
(\ref{dualizequ}) may be very difficult technically. Fortunately,
this problem can be reduced to a purely algebraic one using the
dualisation in the matrix form (\ref{def_N}) as described in the
previous section. Taking into account that  the matrix ${\cal N}$
transforms as
$$
 {\cal N}\ '=g^T{\cal N}(g^T)^{-1}\quad (\hbox{or } {\cal N}\ '=g{\cal N}g^{-1}\
 ),\quad g\in SO(4,4)
$$ and using the relations (\ref{rel_for_N}) one can easily obtain the
desired quantities $(a^p)'$ and $(A^I)'$.

We will denote the 28 generators of the $so(4,4)$ algebra as $${\cal
T}=(H_1,H_2,H_3,H_4,\ P^{\pm I},\, W_{\pm I},\, Z_{\pm
I},\,\Omega^{\pm p},\,X^{\pm}),$$ with $I= 1,2,3,\, p=7,8$. Their
matrix representation   can be found in the Appendix. The
corresponding one-parametric transformations $g=\e^{\alpha{\cal
T}}$, where $\alpha$ is a transformation parameter, give the set of
the target space isometries.
\subsection{Asymptotic conditions}
An important question is how to identify the isometries we need to
use in order to construct solutions with the desired properties.
These are usually associated with asymptotic conditions. In this
paper we consider asymptotic conditions corresponding to 5D
Kaluza-Klein black holes with squashed horizons  embedded into
eleven dimensions which correspond to the following asymptotic
manifold: $T^6\times \mathbb{R}^{1}\times S_{sq}$, where $S_{sq}$ is
a squashed $S^3$. We will assume that target space potentials have
the following general asymptotic behavior
 \be
  \lambda\sim\left(\begin{array}{cc}
  1 & 0 \\
  0 & -1 \\\end{array}\right) +\frac{\delta\lambda}{r},\quad
  \omega_7\sim \frac{\delta\omega_7}{r},\quad \omega_8\sim
 \frac{\delta\omega_8}{r^2}, \quad A_{[3]}=0,\label{asym_cond}
  \ee
 where $\delta\lambda,\ \delta\omega_7$ and $\delta\omega_8$ are
 constant.
The asymptotic behavior with
$\delta\lambda=\delta\omega_7=\delta\omega_8=0$ correspond to the
trivial $S^1$ bundle over a 4D Minkowski space-time. The asymptotic
coset matrix for this case is ${\cal M}_{as}=K$ which is preserved
under the isometries belonging to the isotropy group $H$ of the
$SO(4,4)$:
$$
P^I+ P^{-I},\quad Z_I+ Z_{-I},\quad W_I-W_{-I},\quad X^{+}+
 X^{-},\quad \Omega^7+ \Omega^{-7},\quad
 \Omega^8-\Omega^{-8}.
$$
For more general asymptotic behavior such as (\ref{asym_cond}) one
have use the above transformations with some constraints on the
parameters.

To apply these isometries in the case of minimal 5D supergravity one
needs to find the relevant embedding of the $G_{2(2)}$ subgroup into
 $SO(4,4)$. As was shown in \cite{GS}, the following combinations of the
$SO(4,4)$ generators realize the positive and negative root
generators of $G_{2(2)}$:
$$
 P^{\pm}\sim \sum P^{\pm I},\quad Z_{\pm}\sim \sum Z_{\pm I},\quad
 W_{\pm}\sim \sum W_{\pm I},\quad \Omega^{\pm p},\quad X^{\pm}.
$$
Thus the isometries
$$
P^{+}+ P^{-},\quad Z_{+}+ Z_{-},\quad W_{+}-W_{-},\quad X^{+}+
 X^{-},\quad \Omega^7+ \Omega^{-7},\quad
 \Omega^8-\Omega^{-8}
$$
can be used to generate  new KK solutions in the minimal 5D
supergravity.
\section{Five-parametric squashed black hole}

Our improved generating technique allows us  to construct the
charged Rasheed solution from the Kerr Black Hole. We define the
coordinates  $z^7=x^5,\ z^8=t$ and $x^i=(r,\theta,\phi)$. In this
basis the Kerr solution of the mass $M_K$ and the angular momentum
$J_K=aM_K$ smeared into the fifth dimension reads
$$
 ds^2_5=(dx^5)^2-(1-Z)\Bigl(dt+\frac{aZ\sin^2\theta }{1-Z}d\phi\Bigr)^2+\frac{\rho}{\Delta}dr^2
 +\rho d\theta^2+\frac{\Delta}{1-Z}\sin^2\theta d\phi^2,
$$
where
$$
 \rho=r^2+a^2\cos^2\theta,\quad \Delta=r^2-2M_Kr+a^2,\quad
 Z=\frac{2M_Kr}{\rho}.
$$
The corresponding target space variables are:
 \bea
 &&\lambda_{pq}=\left(\begin {array}{cc}
 1&0\\
 0&Z-1\end {array} \right),\quad \tau=1-Z,\nn\\
 &&\omega_7=0,\quad \omega_8=\frac{2M_K a\cos\theta}{\rho},\quad \Bigl(
 a^7_{\phi}=0,\quad a^8_{\phi}=\frac{aZ\sin^2\theta}{1-Z}\Bigr).\nn
 \eea
 The above definitions of the target space potentials lead to the following
  blocks of the coset matrix
 ${\cal M}$
 $$
  {\cal Q}=\left(\begin {array}{cccc}
 0&0&\frac{2M_K a\cos\theta}{\rho}&0\\
 0&0&0&0\\
 0&0&0&0\\
 0&0&0&0 \end {array}\right),\quad {\cal P}=
 \left(\begin {array}{cccc}
 \frac{1}{Z-1}&0&0&0\\
 0&\frac{1}{Z-1}&0&0\\
 0&0&1&0\\
 0&0&0&1 \end {array}\right).
 $$
One can easily obtain the dual matrix  ${\cal N}$ solving the
Eq.(\ref{def_N}) :
$$
 {\cal N}=\left(\begin {array}{cccccccc}
 -\frac{2M_K\Delta\cos\theta}{\rho(1-Z)}&0&0&0&0&0&\frac{Za\sin^2\theta}{1-Z}&0\\
 0&-\frac{2M_K\Delta\cos\theta}{\rho(1-Z)}&0&0&0&0&0&-\frac{Za\sin^2\theta}{1-Z}\\
 0&0&0&0&0&0&0&0\\
 0&0&0&0&0&0&0&0\\
 0&0&0&0&0&0&0&0\\
 0&0&0&0&0&0&0&0\\
 -\frac{Za(r-2M_k)\sin^2\theta}{r(1-Z)}&0&0&0&0&0&\frac{2M_K\Delta\cos\theta}{\rho(1-Z)}&0\\
 0&\frac{Za(r-2M_k)\sin^2\theta}{r(1-Z)}&0&0&0&0&0&\frac{2M_K\Delta\cos\theta}{\rho(1-Z)}\end {array}\right)d\phi.
$$
To obtain the charged dyon solution from the Kerr one we apply to
the seed coset matrices ${\cal M}$ and ${\cal N}$ the following
sequence of global transformations
$$
 g_1=\e^{\alpha(X^{+}+X^-)}\rightarrow g_2=\e^{\beta(\Omega^7+\Omega^{-7})}
 \rightarrow g_3=\e^{\gamma(\Omega^8-\Omega^{-8})}\rightarrow g_4=\e^{\delta\sum_{I} (Z_I+Z_{-I})}
$$
with the constant parameters $\alpha,\beta,\gamma,\delta$. Here we
assume that the matrices ${\cal M}$ and ${\cal N}$ are transformed
under $g=g_1g_2g_3g_4$ as ${\cal M}'=g{\cal M}g^T$ and ${\cal
N}'=g{\cal N}g^{-1}$ respectively. Then we demand that $g_1g_2g_3$
preserve the $O(\frac{1}{r})$ asymptotic behavior of $a^8_{\phi}$
or, equivalently, the $O(\frac{1}{r^2})$ asymptotic behavior of
$\omega_8$. This give the same relation between three parameters
$\alpha,\beta,\gamma$   as in \cite{rash}:
$$
 \tan2\gamma=\tanh\alpha\sinh\beta.
$$
This constraint ensures  the asymptotic flatness and the absence of
the NUT parameter in the four-dimensional solution. Then extracting
the target space variables from ${\cal M}'$ and ${\cal N}'$ ,
transformed KK one-forms $(a^p)'$ and the five-dimensional one-form
$A'$ one can write the metric and the 3-form field of new solution:
 \bea
ds_{11}^2 &=& \sum_{a,a'} \left( (dz^a)^2 + (dz^{a'})^2
  \right)\nn\\
  &+&f(dt+\Omega')^2+\frac{1}{fD}(dx^5+Wd\phi)^2-D\left(\frac{\rho}{\Delta}dr^2
 +\rho d\theta^2+\frac{\Delta}{1-Z}\sin^2\theta d\phi^2\right),
 \nn\\
 A_{[3]}'&=&\sum_{a,a'}\frac{cs}{D}\Bigl\{(A+B)dt-(sC+cE)dx^5+\Bigl[c(XB-WE)-s(WC+YA)\Bigr]d\phi\Bigr\}\wedge dz^a\wedge dz^{a'},\nn
  \eea
with
 \bea
  &&f=\frac{AB}{D^2},\quad \Omega'=\Omega_5 dx^5+\Omega_\phi d\phi,\quad D=Ac^2+Bs^2\nn\\
  &&\Omega_5=\frac{C}{A}s^3-\frac{E}{B}c^3,\quad
  \Omega_\phi=\frac{WC+YA}{A}s^3+\frac{XB-WE}{B}c^3.\nn
  \eea
The functions $A,B,C,E,X,Y,W$ is given by {\small
 \bea
 A&=&\frac{2M_K^2c_{\beta}^2(c_{\alpha}-p)+2M_K\Bigl(r(p-c_{\alpha}c_{\beta}^2)-as_{\alpha}s_{\beta}c_{\beta}^2\cos\theta\Bigr)-p\rho}{p(\rho-2M_k
r)},\label{def_A}\\
 B&=&\frac{2M_K^2(1+c_{\alpha}p)(p-c_{\alpha}c_{\beta}^2)+2M_K\Bigl(as_{\alpha}s_{\beta}(1+c_{\alpha}^2c_{\beta}^2)\cos{\theta}
-r(c_{\alpha}p^2-c_{\alpha}c_{\beta}^2+p)\Bigr)-p\rho}{p(\rho-2M_k
r)},\\
 E&=&\frac{2M_K\Bigl(M_Ks_{\alpha}(c_{\alpha}c_{\beta}^2-p)+rps_{\alpha}-ac_{\alpha}s_{\beta}\cos\theta\Bigr)}{\rho-2M_Kr},\\
 C&=&-\frac{2M_Kc_{\beta}(M_Ks_{\alpha}s_{\beta}+ap\cos\theta)}{\rho-2M_Kr},\\
 W&=&\frac{2M_Kc_{\beta}\Bigl\{M_K\Bigl(as_{\alpha}\sin^2\theta(p-c_{\alpha}c_{\beta}^2)-2s_{\beta}r\cos\theta\Bigr)
+s_{\beta}(r^2+a^2)\cos\theta-as_{\alpha}pr\sin^2\theta\Bigr\}}{p(\rho-2M_Kr)},\\
X&=&-\frac{2M_Kac_{\beta}\sin^2\theta\Bigl(M_K(c_{\alpha}-p)-rc_{\alpha}\Bigr)}{\rho-2M_Kr},\\
Y&=&\frac{-2M_K\Bigl\{M_K\Bigl(2s_{\alpha}pr\cos\theta-as_{\beta}\sin^2\theta(pc_{\alpha}+1)\Bigr)-s_{\alpha}p\cos\theta
(a^2+r^2)+as_{\beta}r\sin^2\theta\Bigr\}}{\rho-2M_Kr},\label{def_Y}
\eea } where
 $$
 p=\sqrt{c_\alpha^2+s_{\alpha}^2s_{\beta}^2},\quad
c_{\star}=\cosh\star,\ s_{\star}=\sinh\star, \quad
 c=\cosh\delta,\ s=\sinh\delta.
  $$
Our new solution contains five free parameters $M_K, a , \alpha,
\beta, \delta$ and reduces to that of \cite{tym} if $\alpha=0$.
\section{Asymptotic  behavior}
The functions (\ref{def_A})-(\ref{def_Y}) have the following
asymptotic  behavior at spatial infinity
 \bea
A&=&-1-\frac{2M_Kc_{\beta}^2c_{\alpha}}{p}\frac{1}{r}+O\Bigl(\frac{1}{r^2}\Bigr),\nn\\
B&=&1+\frac{2M_Kc_{\alpha}(c_{\beta}^2-p^2)}{p}\frac{1}{r}+O\Bigl(\frac{1}{r^2}\Bigr),\nn\\
C&=&-{2M_Kc_{\beta}(s_{\alpha}s_{\beta}M_K+pa\cos{\theta})}\frac{1}{r^2}+O\Bigl(\frac{1}{r^3}\Bigr),\nn\\
E&=&2M_Ks_{\alpha} p\frac{1}{r}+O\Bigl(\frac{1}{r^2}\Bigr),\nn\\
X&=&2M_Kac_{\alpha}c_{\beta}\sin^2{\theta}\frac{1}{r}+O\Bigl(\frac{1}{r^2}\Bigr),\nn\\
Y&=&2M_ks_{\alpha}p\cos\theta-2M_Kas_{\beta}\sin^2\theta\frac{1}{r}+O\Bigl(\frac{1}{r^2}\Bigr),\nn\\
W&=&\frac{2M_Ks_{\beta}c_{\beta}\cos\theta}{p}-2M_Kac_{\beta}s_{\alpha}\sin^2\theta\frac{1}{r}+O\Bigl(\frac{1}{r^2}\Bigr),\nn
 \eea
These decompositions lead to the asymptotical expression of the
five-dimensional metric and the electro-magnetic one-form $A'$:
 \bea
 ds_5^2&=&-dt^2+(dx^5+\frac{2M_Ks_{\beta}c_{\beta}\cos\theta}{p}d\phi)^2+dr^2
 +r^2 (d\theta^2+\sin^2\theta d\phi^2),\nn\\
 A_{[3]}'&=&\sum_{a,a'}A'\wedge dz^a\wedge dz^{a'},\nn\\
 A_{t}'&=&\frac{2M_Kcspc_{\alpha}}{r}+O\Bigl(\frac{1}{r^2}\Bigr),\quad
 A_{x^5}'=\frac{2M_Kc^2sps_{\alpha}}{r}+O\Bigl(\frac{1}{r^2}\Bigr),\nn\\
 A_{\phi}'&=&2M_Kcs\Bigl\{-sps_{\alpha}\cos\theta+\frac{a\sin^2\theta(ss_{\beta}-cc_{\alpha}c_{\beta})
 +2M_Ks_{\alpha}\cos\theta(cc_{\beta}s_{\beta}+s^3p^2c_{\alpha})}{r}\Bigr\}+O\Bigl(\frac{1}{r^2}\Bigr),\nn
  \eea
Then we define the Komar mass and angular momenta as
 \bea
 &&M=\frac{1}{2\pi^2}\int{d\Sigma_{\alpha\beta}}\xi_{(t)}^{\alpha;\beta},\nn\\
 &&J_{\phi}=-\frac{1}{2\pi^2}\int{d\Sigma_{\alpha\beta}}\xi_{(\phi)}^{\alpha;\beta},\nn\\
 &&J_{x^5}=-\frac{1}{2\pi^2}\int{d\Sigma_{\alpha\beta}}\xi_{(x^5)}^{\alpha;\beta},\nn
 \eea
 where $\xi_{(t)}^{\alpha},\ \xi_{(\phi)}^{\alpha},\ \xi_{(x^5)}^{\alpha}$ are the Killing vector
 fields
$\xi_{(t)}=\xi_{(t)}^{\alpha}\partial_{\alpha}=\partial_t,\
\xi_{(\phi)}=\partial_{\phi},\ \xi_{(x^5)}=\partial_{x^5},$ which
normalized as $\xi_{(t)}^2=-1,\ \xi_{(\phi)}^2=1,\ \xi_{(x^5)}^2=1$
at infinity. The integrals are taken over the squashed $S^3$ at
spatial infinity $r \rightarrow \infty$ and the surface element is
$d\Sigma_{tr}=r^2\sin(\theta)d\theta\wedge d\phi\wedge dx^5$. Here
we assume that $x^5\in S^1$ has the periodicity $2\pi R_5$. The
computations of the Komar integrals with respect the 5-dimensional
metric
$$
ds_5^2=f(dt+\Omega')^2+\frac{1}{fD}(dx^5+Wd\phi)^2-D\left(\frac{\rho}{\Delta}dr^2
 +\rho d\theta^2+\frac{\Delta}{1-Z}\sin^2\theta d\phi^2\right)
$$
give the following results:
 \bea
 &&M=8R_5(c^2+s^2)M_Kpc_{\alpha},\nn\\
 &&J_{\phi}=-\frac43
 R_5M_Ka(c_{\alpha}c_{\beta}c^3-s_{\beta}s^3),\nn\\
 &&J_{x^5}=-4R_5M_Kps_{\alpha}c^3\nn
 \eea
 The conserved electric  charge $Q_e$ of the new solution is
 $$
  Q_e=-\frac{1}{4\pi G_5}\int d\Sigma_{\alpha\beta}(F^{'\alpha\beta}+
  \frac{1}{\sqrt3\sqrt{-g}}\varepsilon^{\alpha\beta\gamma\delta\eta}A_{\gamma}'F_{\delta\eta}'),
 $$
where $F'=dA'$. One finds that
$$
 Q_e=\frac{8\pi}{G_5}R_5M_Kscc_{\alpha}p
$$

\section{Conclusions}
We have presented a new formulation of  solution generating
technique for the 5D minimal and $\ U(1)^3$ supergravities based on
the 3D sigma-model with the $SO(4,4)$ isometry group. Starting  from
any seed solution possessing two commuting Killing vector fields and
using transformations of the target space isometry group one can
construct new solutions with the same three-dimensional metric. The
solution generation procedure consists in solving the dualisation
equations for the seed solution to express it in the sigma-model
variables, applying some $SO(4,4)$ transformations to get new
sigma-model potential, and finally to pass back to the metric and
field variable. Usually the last steps also involves solving the
dualisation equations, but we suggest here the dualisation in the
matrix form with an independent transformation of the dual
variables. This allows to avoid solving differential equations for
the backward dualisation, replacing this step by an algebraic
procedure. As an application we have obtained the five-parametric
Kaluza-Klein black hole of the minimal 5D supergravity. Our
generating transformations generalize those of the vacuum 5D gravity
to the presence of vector fields and open a way to develop the
inverse scattering technique for this more general case.

 \begin{acknowledgments}
The authors are grateful to Gerard Cl\'ement and Chiang-Mei-Chen for
helpful discussions. The paper was supported by the RFBR grant
08-02-01398-a.
 \end{acknowledgments}

\appendix

\section{$8\times 8$ matrix representation }
We choose the following $8\times 8$ matrix representation of the
so(4,4) algebra
 \be
  E= \left( \begin{array}{cc}
A & B \\
C & -A^{\widehat{T}} \end{array} \right),\label{def_matr_basis}
 \ee
 where
$A,\ B,\ C$ are the $4\times 4$ matrices, $A,\ B$ being
antisymmetric, $B=-B^{T},\ C=-C^{T}$, and the symbol $\widehat{T}$
in $A^{\widehat{T}}$ means  transposition with respect to the minor
diagonal. The diagonal matrices $\vec H$ are given by the following
$A-$type matrices (with $B=0=C$): \bea
 A_{H_1}\!\!&=&\!\!\left( \begin {array}{cccc}
 \sqrt{2}&0&0&0\\0&0&0&0\\0&0&0&0
 \\0&0&0&0\end {array} \right), \
 A_{H_2}\!=\!\left(\begin {array}{cccc} 0&0&0&0\\0&\sqrt{2}&0&0\\0&0&0&0
 \\0&0&0&0\end {array} \right),\
 A_{H_3}\!=\!\left(\begin {array}{cccc} 0&0&0&0\\0&0&0&0\\0&0&\sqrt{2}&0
 \\0&0&0&0\end {array} \right),\
 A_{H_4}\!=\!\left(\begin {array}{cccc} 0&0&0&0\\0&0&0&0\\0&0&0&0
 \\0&0&0&\sqrt{2}\end {array} \right).\nn
 \eea
Twelve generators corresponding to the positive roots are given by
the upper-triangular matrices $E_k,\ k=1,\ldots,12, $.  From these
the generators labeled by $k=2,4,6,7,9,12$ are of pure $A$-type
(with $B=0=C$):
 \bea
 &&A_{E_2}=\left(\begin {array}{cccc} 0&0&0&1\\0&0&0&0\\0&0&0&0
 \\0&0&0&0\end {array} \right),\quad
 A_{E_4}=\left(\begin {array}{cccc} 0&0&0&0\\0&0&0&0\\0&0&0&-1
 \\0&0&0&0\end {array} \right), \quad
 A_{E_6}=\left(\begin {array}{cccc} 0&1&0&0\\0&0&0&0\\0&0&0&0
 \\0&0&0&0\end {array} \right),\nn\\
 &&A_{E_7}=\left(\begin {array}{cccc} 0&0&0&0\\0&0&0&-1\\0&0&0&0
 \\0&0&0&0\end {array} \right)\quad
 A_{E_9}=\left(\begin {array}{cccc} 0&0&-1&0\\0&0&0&0\\0&0&0&0
 \\0&0&0&0\end {array} \right),\quad
A_{E_{12}}=\left(\begin {array}{cccc} 0&0&0&0\\0&0&-1&0\\0&0&0&0
 \\0&0&0&0\end {array} \right).\nn\eea
 while the other six are of pure $B$ type (with $A=0=C$):
 \bea &&B_{E_1}=\left(\begin {array}{cccc} 1&0&0&0\\0&0&0&0\\0&0&0&0
 \\0&0&0&-1\end {array} \right),\quad
 B_{E_3}=\left(\begin {array}{cccc} 0&0&0&0\\0&-1&0&0\\0&0&1&0
 \\0&0&0&0\end {array} \right),\quad
 B_{E_5}=\left(\begin {array}{cccc} 0&0&0&0\\0&0&0&0\\-1&0&0&0
 \\0&1&0&0\end {array} \right),\nn
 \eea
 \bea
 &&B_{E_8}=\left(\begin {array}{cccc} 0&0&0&0\\-1&0&0&0\\0&0&0&0
 \\0&0&1&0\end {array} \right),\quad
 B_{E_{10}}=\left(\begin {array}{cccc} 0&1&0&0\\0&0&0&0\\0&0&0&-1
 \\0&0&0&0\end {array} \right),\quad
 B_{E_{11}}=\left(\begin {array}{cccc} 0&0&1&0\\0&0&0&-1\\0&0&0&0
 \\0&0&0&0\end {array} \right).\nn
 \eea
The correspondence with the previously introduced generators is as
follows ($I=1,2,3,\;p=7,8$): \be P^I\leftrightarrow {E_I}, \quad
W_I\leftrightarrow { E_{I+3}}, \quad Z_I\leftrightarrow {E_{I+6}},
\quad \Omega^p\leftrightarrow {E_{p+3}},\quad X^{+} \leftrightarrow
{E_{12}}.\nn\ee In this representation, the matrices corresponding
to the negative roots, \be P^{-I}\leftrightarrow E_{-I}, \quad
W_{-I}\leftrightarrow E_{-(I+3)}, \quad Z_{-I}\leftrightarrow
E_{-(I+6)}, \quad \Omega^{-p}\leftrightarrow E_{-(p+3)},\quad X^{-}
\leftrightarrow E_{-12},\nn\ee are  transposed with respect to the
positive roots matrices:  \be
 E_{-k}=(E_k)^T.\nn
 \ee
The following normalization conditions are assumed: \be
 \mathrm{Tr}(H_i,H_j)=4\delta_{ij},\ i,j=1\ldots 4,\qquad
 \mathrm{Tr}(E_k,E_{-k})=2,\nn
 \ee
and the involution matrix $K$   is chosen as \be
 K=\rm{diag}(\kappa,\kappa,1,1,1,1,\kappa,\kappa).\nn
 \ee
The generators of the isotropy subgroup are selected by the equation
$h(\Phi)^T K h(\Phi)=K$. They are given by the following linear
combinations of the generators: \be
 P^I-\kappa P^{-I},\quad Z_I -\kappa Z_{-I},\quad W_I-W_{-I},\quad
 X^{+}-\kappa X^{-},\quad \Omega^7-\kappa \Omega^{-7},\quad
 \Omega^8-\Omega^{-8}.
\nn \ee
\section{Matrix representation of coset ${\cal M}$}
$$
 {\cal M}=\left( \begin {array}{cc}
 {\cal P}&{\cal P}{\cal Q}\\
 {\cal Q}^T{\cal P}&\widetilde{\cal P}+{\cal Q}^T{\cal P}{\cal Q}
 \end {array} \right),
 $$
 where the $4\times 4$ blocks ${\cal P}$ and ${\cal Q}$ are
\bea
 &&{\cal Q}=\left( \begin{array}{cccc}
 \mu_1+ \frac{u^3v^2-v^3u^2}{2}&\omega_7-
 \frac{u^3v^1u^2-2u^3v^2u^1+v^3u^1u^2}{6}-u^2\mu_2,&
 \omega_8+\frac{v^3u^1v^2-2v^3u^2v^1+u^3v^1v^2}{6}-v^2\mu_2&0\\
  -v^2&-\mu_3+\frac{v^1u^2-u^1v^2}{2}&0&\\
   -u^2&0&&\\
    0&&&\\
 \end{array} \right),\nn\\
 &&{\cal P}=\left( \begin {array}{cc}
 \Psi^T\Lambda \Psi,&\Psi^T\Lambda\Phi\\
 \Phi^T\Lambda\Psi,&\Phi^T\Lambda\Phi+\e^{\sqrt2\phi_4}
 \end {array} \right),\label{def_P}\quad \widetilde{\cal P}=({\cal
 P}^{-1})^{\widehat T}.\nn
 \eea
 Here $\Psi$ and $\Lambda$ are the $3\times 3$ matrices
 $$
 \Psi=\left( \begin {array}{ccc}
 1&u^3&-v^3\\
 0&1&0 \\
 0&0&1 \end {array}
 \right),\quad
 \Lambda=\kappa\left( \begin {array}{ccc}
 \e^{\sqrt2\phi_1}&0&0\\
 0&\e^{\sqrt2\phi_2}&-\chi\e^{\sqrt2\phi_2} \\
 0&-\chi\e^{\sqrt2\phi_2}&\e^{\sqrt2\phi_2}\chi^2+\kappa\e^{\sqrt2\phi_3} \end {array}
 \right)
  $$
  and $\Phi$ is the 3-column
  $$
   \Phi=\left( \begin {array}{ccc}
 \mu_2+\frac12(u^1v^3-u^3v^1)\\
 -v^1\\
 -u^1\end {array}
 \right).
  $$


\begin{thebibliography}{20}

\bibitem{empare} R. Emparan and H.S. Reall,Class. Quant. Grav. \textbf{23},
R169 (2006) [arXiv:hep-th/0608012]; Living Rev. Rel. {\bf{11}}, 6
(2008) [arXiv:0801.3471].

\bibitem{gerard} Gerard Clement,
``Sigma-model approaches to exact solutions in higher-dimensional
gravity and supergravity'', Talk presented at the WE Heraeus Seminar
on Models of Gravity in Higher Dimensions: From Theory to
Experimental Search, Bremen, 25-29.8.2008 [arXiv:0811.0691].

\bibitem{ga} D.V. Gal'tsov, ``Integrable systems in stringy gravity'',
Phys. Rev. Lett. {\bf{74}}, 2863 (1995) [arXiv:hep-th/9410217].

\bibitem{BeZa}
  V.~A.~Belinsky and V.~E.~Zakharov,
  ``Integration Of The Einstein Equations By The Inverse Scattering
Problem
  Technique And The Calculation Of The Exact Soliton Solutions,''
  Sov.\ Phys.\ JETP {\bf 48}, 985 (1978)
  [Zh.\ Eksp.\ Teor.\ Fiz.\  {\bf 75}, 1953 (1978)].
  V.~A.~Belinsky and V.~E.~Zakharov,
  ``Stationary Gravitational Solitons With Axial Symmetry,''
  Sov.\ Phys.\ JETP {\bf 50}, 1 (1979)
  [Zh.\ Eksp.\ Teor.\ Fiz.\  {\bf 77}, 3 (1979)].

\bibitem{Har}
  T.~Harmark,
  ``Stationary and axisymmetric solutions of higher-dimensional general
  relativity,'' Phys.\ Rev.\  D {\bf 70}, 124002 (2004),
  [arXiv:hep-th/0408141];
R.~Emparan, T.~Harmark, V.~Niarchos, N.~A.~Obers and
M.~J.~Rodriguez,
  ``The Phase Structure of Higher-Dimensional Black Rings and Black
Holes,'' JHEP {\bf 0710}, 110 (2007)
  [arXiv:0708.2181 [hep-th]];
\bibitem{inve}
  H.~Elvang and P.~Figueras,
  ``Black Saturn,''
  JHEP {\bf 0705}, 050 (2007)
  [arXiv:hep-th/0701035];
  H.~Iguchi and T.~Mishima,
  ``Black di-ring and infinite nonuniqueness,''
  Phys.\ Rev.\  D {\bf 75}, 064018 (2007)
  [arXiv:hep-th/0701043];
  J.~Evslin and C.~Krishnan,
  ``The Black Di-Ring: An Inverse Scattering Construction,''
  arXiv:0706.1231 [hep-th];
  H.~Elvang and M.~J.~Rodriguez,
  ``Bicycling Black Rings,''
  arXiv:0712.2425 [hep-th];
  J.~Evslin and C.~Krishnan,
  ``Metastable Black Saturns,''
  arXiv:0804.4575 [hep-th].
  J.~Evslin,
  JHEP {\bf 0809}, 004 (2008)
  [arXiv:0806.3389 [hep-th]];
  J.~Evslin and C.~Krishnan,
  arXiv:0706.1231 [hep-th].
  H.~Iguchi and T.~Mishima,
  ``Solitonic generation of five-dimensional black ring solution,''
  Phys.\ Rev.\  D {\bf 73}, 121501 (2006)
  [arXiv:hep-th/0604050];
  S.~Tomizawa, H.~Iguchi and T.~Mishima,
  ``Relationship between solitonic solutions of five-dimensional Einstein
  equations,''
  Phys.\ Rev.\  D {\bf 74}, 104004 (2006)
  [arXiv:hep-th/0608169];
  S.~Tomizawa and M.~Nozawa,
  ``Vaccum solutions of five-dimensional Einstein equations generated by
  inverse scattering method. II: Production of black ring solution,''
  Phys.\ Rev.\  D {\bf 73}, 124034 (2006)
  [arXiv:hep-th/0604067];
  T.~Koikawa,
  ``Infinite number of soliton solutions to 5-dimensional vacuum Einstein
  equation,''
  Prog.\ Theor.\ Phys.\  {\bf 114}, 793 (2005)
  [arXiv:hep-th/0501248];
  T.~Azuma and T.~Koikawa,
  ``Infinite number of stationary soliton solutions to five-dimensional vacuum
  Einstein equation,''
  Prog.\ Theor.\ Phys.\  {\bf 116}, 319 (2006)
  [arXiv:hep-th/0512350];
  T.~Mishima and H.~Iguchi,
  ``New axisymmetric stationary solutions of five-dimensional vacuum  Einstein
  equations with asymptotic flatness,''
  Phys.\ Rev.\  D {\bf 73}, 044030 (2006)
  [arXiv:hep-th/0504018];
  S.~Tomizawa, Y.~Morisawa and Y.~Yasui,
  ``Vacuum solutions of five dimensional Einstein equations generated by
  inverse scattering method,''
  Phys.\ Rev.\  D {\bf 73}, 064009 (2006)
  [arXiv:hep-th/0512252];

\bibitem{yaza}
  S.~S.~Yazadjiev,
  ``Magnetized static black Saturn,''
  arXiv:0802.0784 [hep-th];
  S.~S.~Yazadjiev,
  ``Completely integrable sector in 5D Einstein-Maxwell gravity and  derivation
  of the dipole black ring solutions,''
  [arXiv:hep-th/0602116];
  S.~S.~Yazadjiev,
  ``Solution generating in 5D Einstein-Maxwell-dilaton gravity and derivation
  of dipole black ring solutions,''
  [arXiv:hep-th/0604140];
  S.~S.~Yazadjiev,
  ``Black Saturn with dipole ring,''
  [arXiv:0705.1840 [hep-th]].

\bibitem{bccgsw} A.~Bouchareb, G.~Cl\'ement, C-M.~Chen,
D.~V.~Gal'tsov, N.~G.~Scherbluk, and Th. Wolf, ``G2 generating
technique for minimal $5D$ supergravity and black rings'', Phys.
Rev. D {\bf 76}, 104032 (2007) [arXiv:0708.2361].

\bibitem{clem} G. Cl\'ement, Journ. Math. Phys. \textbf{49}, 042503 (2008);
Erratum, Journ. Math. Phys. \textbf{49}, 079901 (2008)
[arXiv:0710.1192].

\bibitem{mizo}
  S.~Mizoguchi and N.~Ohta,
  ``More on the similarity between D = 5 simple supergravity and M theory,''
  Phys.\ Lett.\  B {\bf 441}, 123 (1998)
  [arXiv:hep-th/9807111].
  E.~Cremmer, B.~Julia, H.~L\"u and C.~N.~Pope,
  ``Higher-dimensional origin of D = 3 coset symmetries,''
  arXiv:hep-th/9909099.
  S.~Mizoguchi and G.~Schr\"{o}der,
  ``On discrete U-duality in M-theory,''
  Class.\ Quant.\ Grav.\  {\bf 17}, 835 (2000)
  [arXiv:hep-th/9909150].
  M.~Possel,
  ``Hidden symmetries in five-dimensional supergravity,''
   PhD Thesis, Hamburg, 2003.
  M.~Possel and S.~Silva,
  ``Hidden symmetries in minimal five-dimensional supergravity,''
  Phys.\ Lett.\  B {\bf 580}, 273 (2004)
  [arXiv:hep-th/0310256].
   M. Gunaydin, A. Neitzke, O. Pavlyk and B. Pioline, "Quasi-conformal
actions, quaternionic discrete series and twistors: $SU(2,1)$ and
$G_2(2)$" [arXiv:0707.1669]

\bibitem{gugu}
  M.~Gunaydin and F.~Gursey,
  ``Quark structure and octonions,''
  J.\ Math.\ Phys.\  {\bf 14}, 1651 (1973).

\bibitem{GS} D.V. Gal'tsov and N.G. Scherbluk,
``Generating technique for $U(1)^3 5D$ supergravity,'' Phys. Rev. D
\textbf{78}, 064033 (2008) [arXiv:0805.3924].

\bibitem{posen}
  A.~A.~Pomeransky and R.~A.~Sen'kov,
  ``Black ring with two angular momenta,''
  arXiv:hep-th/0612005.


\bibitem{IM}
 H.~Ishihara and K.~Matsuno, `` Kaluza-Klein black holes with squashed horizons,''
 Prog.\ Theor.\ Phys. {\bf 116} 417, (2006)
 [arXiv: hep-th/0510094].

\bibitem{Ca}
  R.~G.~Cai, L.~M.~Cao and N.~Ohta,
  Phys.\ Lett.\  B {\bf 639}, 354 (2006)
  [arXiv:hep-th/0603197].
\bibitem{Kuri}
  Y.~Kurita and H.~Ishihara,
  ``Mass and Free energy in Thermodynamics of Squashed Kaluza-Klein Black
  Holes,''
  Class.\ Quant.\ Grav.\  {\bf 24}, 4525 (2007)
  [arXiv:0705.0307 [hep-th]];
   ``Thermodynamics of Squashed Kaluza-Klein Black Holes and Black Strings -- A
   Comparison of Reference Backgrounds,''
  Class.\ Quant.\ Grav.\  {\bf 25}, 085006 (2008)
  [arXiv:0801.2842 [hep-th]].

\bibitem{wang}
 T.~Wang, ``A Rotating Kaluza-Klein black hole with squashed horizons,''
 Nucl.\ Phys. \ B {\bf 756} 86-99, (2006)
 [arXiv: hep-th/0605048].

\bibitem{NIMT}
 T.~Nakagawa, H.~Ishihara, K.~Matsuno, and S.~Tomizawa,
 ``Charged Rotating Kaluza-Klein Black Holes in Five Dimensions,''
 Phys.\ Rev. \ D {\bf 77},044040, (2008)
 [arXiv:0801.0164].


\bibitem{TIMN}
 S.~Tomizawa, H.~Ishihara, K.~Matsuno, and T.~Nakagawa,
 ``Squashed Kerr-Godel Black Holes: Kaluza-Klein Black Holes with Rotations of Black Hole and Universe,''
  [arXiv:0803.3873]. (2008)

\bibitem{MINT}
  K.~Matsuno, H.~Ishihara, T.~Nakagawa, and S.~Tomizawa,
 ``Rotating Kaluza-Klein Multi-Black Holes with Godel Parameter,''
 Phys.\ Rev. \ D {\bf 78},064016, (2008)
 [arXiv:0806.3316].

\bibitem{TI}
  S.~Tomizawa and A.~Ishibashi,
  ``Charged Black Holes in a Rotating Gross-Perry-Sorkin Monopole Background,''
  [arXiv:0807.1564]. (2008)
\bibitem{tym}
 S.~Tomizawa, Y.~Yasui, and Y.~Morisawa, ``Charged Rotating Kaluza-Klein Black Holes Generated by G2(2) Transformation,''
 [arXiv:0809.2001]. (2008)

\bibitem{rash}
 D.~Rasheed, ``The rotating dyonic black holes of Kaluza-Klein
 theory,'' Nucl.\ Phys.\ B {\bf 454} 379-401, (1995)
 [arXiv:hep-th/9505038].
 \bibitem{Fre} P.~Fre ``Lectures on Special K\"{a}hler Geometry and Electric–Magnetic
Duality Rotations,'' Nucl. Phys. B (Proc. Suppl.) {\bf 45B,C} (1996)
59-114 [arXiv:hep-th/9512043]

\bibitem{BMG} P.~Breitenlohner, D.~Maison, and G.~Gibbons,
``4-Dimensional Black Holes from Kaluza-Klein theories,'' Commun.
Math. Phys. {\bf 120} 295–334 (1987)

\bibitem{BM} P.~Breitenlohner, and D.~Maison,
 ``On nonlinear sigma models arising in (super-)gravity,''
Commun.Math.Phys. {\bf 209}:785-810, (2000) [arXiv:gr-qc/9806002]

\bibitem{abcaffm}
L.~Andrianopoli, M.~Bertolini, Anna Ceresole, R.~D'Auria,
S.~Ferrara, P.~Fre, T.~Magri ``N=2 supergravity and N=2
superYang-Mills theory on general scalar manifolds: Symplectic
covariance, gaugings and the momentum map,'' J. Geom. Phys. {\bf 23}
111-189, (1997) [arXiv:hep-th/9605032]

\end{thebibliography}
\end{document}